\documentclass[letterpaper,twocolumn,aps, reprint, pra]{revtex4-2}
\usepackage{mathptmx}

\usepackage[T1]{fontenc}
\usepackage[latin9]{inputenc}
\usepackage{color}
\usepackage{units}
\usepackage{amsmath}
\usepackage{graphicx}
\usepackage[pdfusetitle,
 bookmarks=true,bookmarksnumbered=false,bookmarksopen=false,
 breaklinks=false,pdfborder={0 0 0},pdfborderstyle={},backref=false,colorlinks=true]
 {hyperref}

\makeatletter

\pdfpageheight\paperheight
\pdfpagewidth\paperwidth

\makeatother

\begin{document}
\title{Bath-free squeezed phonon lasing via intrinsic ion-phonon coupling}
\author{Chen-Yu Lee$^{1}$ and Guin-Dar Lin$^{1,2,3}$}
\affiliation{$^{1}$Physics Division, National Center for Theoretical Sciences,
Taipei 10617, Taiwan~\linebreak{}
$^{2}$Department of Physics and Center for Quantum Science and Engineering,
National Taiwan University, Taipei 10617, Taiwan~\linebreak{}
$^{3}$Trapped-Ion Quantum Computing Laboratory, Hon Hai Research
Institute, Taipei 11492, Taiwan}
\begin{abstract}
We present a theoretical model for realizing squeezed lasing in a
trapped-ion system without relying on engineered baths or tailored
dissipative reservoirs.\textcolor{blue}{{} }Our approach leverages the
intrinsic ion-phonon interactions, where two trapped ions, each interacting
with a shared vibrational mode, are driven on both red- and blue-sideband
transitions. This enables the creation of a squeezed state of motion
through the dynamic coupling between the ions\textquoteright{} internal
states and the phonon mode. Unlike traditional methods that require
bath engineering, our model demonstrates that squeezed lasing can
be achieved through a direct manipulation of ion-phonon interactions,
with no external reservoirs required. We explore the steady-state
behavior of the system, analyzing the onset of lasing, gain-loss balance,
and the role of the squeezing parameter in shaping the phonon field's
statistical properties. Furthermore, we show how external coherent
drives can stabilize phase coherence and achieve controlled quadrature
squeezing, offering a simple yet effective method for achieving squeezed
lasing in quantum mechanical systems. Our findings provide new insights
into the realization of squeezed states in phonon-based systems, with
potential applications in quantum metrology and information processing.
\end{abstract}
\maketitle
\hypersetup{
colorlinks=true,
urlcolor=blue,
linkcolor=blue,
citecolor=blue
}

\section{Introduction}

Laser technology plays an indispensable role in various fields, from
scientific research to industrial applications and consumer electronics.
The distinctive features of optical lasers, including their quantum
coherence and long-range propagation, have made them vital tools in
modern quantum engineering and communication. In recent years, there
has been a growing interest in the acoustic counterpart of lasing
phenomena, specifically phonon lasing. This research extends the framework
of quantum optics to explore mechanical degrees of freedom, which
share fundamental mathematical similarities with photonic systems.
Phonons, with their slower dynamics compared to photons, allow for
more controlled manipulation of system parameters, enabling precise
phase control within a given period. Furthermore, unlike photons,
phonons can be coupled to atomic systems deterministically through
state-dependent kicks, as seen in trapped-ion quantum computing and
simulation \citep{Cirac1995}, offering an advantage over the probabilistic
photon-atom interaction.

Phonon lasers have emerged as a promising platform for quantum technologies,
with potential applications in quantum computing \citep{Soderberg2010,Ruskov2012,Reinke2016,Fluehmann2019,Ge2019,Loye2020,Gan2020,Nguyen2021,Chen2021,Wan2021},
communications \citep{Bienfait2019,Dumur2021}, and metrology \citep{Dalvit2006,Sabin2014,Hu2018,Wolf2019,Drechsler2020,Cerrillo2021,Delakouras2022}.
The first demonstration of a phonon laser in a trapped-ion system
driven by optical forces \citep{Vahala2009} marked a significant
milestone, showing self-sustained oscillations beyond a certain threshold
in a single ion, driven by optical energy. Following this breakthrough,
numerous proposals and experimental demonstrations have been conducted
in a variety of platforms, such as trapped ions \citep{Knuenz2010,Ip2018,Behrle2023},
quantum dots \citep{Kabuss2012,Kabuss2013,Khaetskii2013}, optomechanical
systems \citep{Grudinin2010,Beardsley2010,Khaetskii2013,Mahboob2013,Kemiktarak2014,Jing2014,Zhang2018,Jiang2018,Pettit2019,Sheng2020}.

A variety of important laser-like properties have been observed in
phonon systems, including threshold behavior \citep{Vahala2009,Grudinin2010,Khurgin2012,Mahboob2013,Kemiktarak2014,Zhang2018,Pettit2019,Sheng2020},
Poissonian statistics \citep{Pettit2019}, linewidth narrowing \citep{Grudinin2010,Beardsley2010,Khurgin2012,Mahboob2013,Zhang2018,Pettit2019},
injection locking \citep{Knuenz2010,Ip2018}, mode competition \citep{Kemiktarak2014,Sheng2020},
and phase control \citep{Zhang2018a}. While these methods are effective
in realizing lasing behavior in certain systems, they often lack flexibility,
as key system parameters, such as the lasing resonator or dissipation
pathways, are fixed once designed, making in-situ reconfiguration
challenging.

On the other hand, squeezed laser systems in optical setups have typically
relied on bath engineering to induce squeezing, where engineered dissipative
environments are used to tailor the quantum statistics of the emitted
field \citep{NavarreteBenlloch2014,SanchezMunoz2021}. Here, we take
a different route: squeezing arises from coherent ion-phonon interactions
and can be absorbed into a redefinition of the motional mode, without
altering the underlying gain and loss mechanisms. This separation
allows squeezed lasing to be realized without bath engineering, while
preserving the standard laser threshold physics.

We implement this idea in a minimal trapped-ion setting, where two
ions are coupled to a shared vibrational mode and simultaneously driven
on their red- and blue-sideband transitions. These coherent drives
generate effective ion-phonon interactions that realize squeezed dynamics
directly at the level of the motional mode, without introducing additional
dissipative channels. The resulting scheme remains experimentally
simple, while providing a controllable platform for exploring squeezed-lasing
physics in isolated mechanical systems.

This paper is organized as follows. In Sec.~II, we introduce the
theoretical model and show how the driven ion-phonon interactions
can be mapped onto a squeezed-mode laser description. In Sec.~III,
we analyze the signatures of squeezed lasing, including threshold
behavior, phonon statistics, and linewidth narrowing. In Sec.~IV,
we study phase symmetry breaking and demonstrate how an external coherent
drive enables controlled quadrature squeezing. In Sec.~V, we discuss
the experimental feasibility of our scheme in current trapped-ion
platforms. Finally, Sec.~VI summarizes our results and outlines possible
extensions.

\section{Theoretical Model}

\begin{figure}[t]
\begin{centering}
\includegraphics[width=8.8cm]{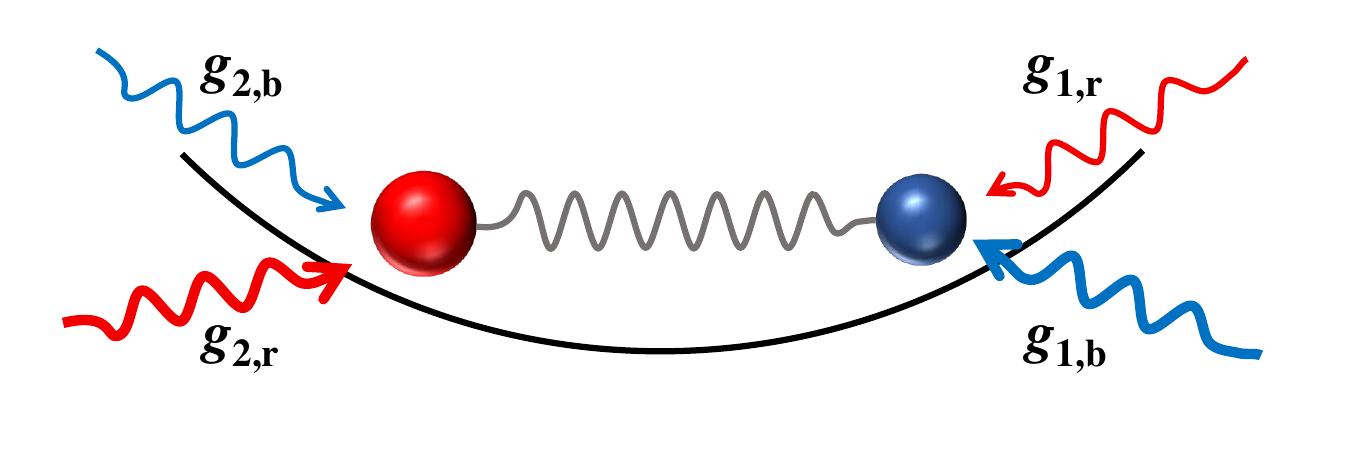}
\par\end{centering}
\caption{Schematic of the bath-free squeezed phonon laser configuration. Two
ions are driven asymmetrically on their red- and blue-sideband transitions.
One ion acts as a heating agent to provide gain, while the other acts
as a cooling agent to introduce loss for the shared motional mode.
This setup generalizes the standard phonon laser architecture to the
squeezed domain by leveraging intrinsic ion-phonon interactions without
the need for external reservoir engineering. \label{fig:model}}
\end{figure}

We consider a physical system comprising two trapped ions coupled
to a shared quantized vibrational mode of the ion crystal, representing
a common axial motional degree of freedom. The vibrational mode is
described by the bosonic annihilation and creation operators $a$
and $a^{\dag},$ respectively. As illustrated in the configuration
in Fig.~\ref{fig:model}, the two ions are assigned distinct functional
roles: one ion serves as a heating agent to provide gain, while the
other facilitates cooling to introduce controlled loss.

To realize the proposed squeezed-lasing dynamics, both ions are simultaneously
driven on their respective red- and blue-motional sidebands. This
dual-tone driving scheme enables coherent Jaynes-Cummings (JC) and
anti-Jaynes-Cummings (AJC) interactions between the ions' internal
electronic states and the shared phonon mode. Under the rotating-wave
approximation and within the Lamb-Dicke regime, the total interaction
Hamiltonian is given by

\begin{align}
H & =\sum_{i=1,2}\left(g_{i,\text{b}}a^{\dagger}\sigma_{i}^{+}+g_{i,\text{b}}^{*}a\sigma_{i}^{-}+g_{i,\text{r}}a\sigma_{i}^{+}+g_{i,\text{r}}^{*}a^{\dagger}\sigma_{i}^{-}\right)\label{eq:Hamiltonain}
\end{align}
where $\sigma_{i}^{+}$ and $\sigma_{i}^{-}$ are the raising and
lowering operators for the internal states of ion $i=1,2.$ The coupling
strengths $g_{i,\text{b/r}}=\left|g_{i,\text{b/r}}\right|e^{-i\phi_{i,\text{b/r}}}$
describe the effective interaction strengths for blue-sideband and
red-sideband transitions, respectively, with $\phi_{i,\text{b/r}}$
the associated phases.

The physical mechanism for achieving squeezing in this model relies
on the mapping of the system onto a standard laser framework within
a squeezed basis. To identify the intrinsic squeezing dynamics, we
define the unitary squeezing transformation
\begin{equation}
S\left(r,\theta\right)=\exp\left[\frac{r}{2}\left(e^{-i\theta}a^{2}-e^{i\theta}a^{\dag2}\right)\right],\label{eq:squeezingt}
\end{equation}
where $r$ is the squeezing parameter and $\theta$ is the squeezing
phase. By appropriately choosing the drive parameters such that $\tanh\left(r\right)=\left|\frac{g_{1,\text{r}}}{g_{1,\text{b}}}\right|=\left|\frac{g_{2,\text{b}}}{g_{2,\text{r}}}\right|<1,$
and aligning the optical phases $\theta=\phi_{1,\text{r}}-\phi_{1,\text{b}}=\phi_{2,\text{r}}-\phi_{2,\text{b}},$
the Hamiltonian in Eq.~\ref{eq:Hamiltonain} can be reparameterized
as

\begin{align}
H & =g_{1}\cosh\left(r\right)a^{\dagger}\sigma_{1}^{+}+g_{1}\sinh\left(r\right)a\sigma_{1}^{+}\nonumber \\
 & +g_{2}\sinh\left(r\right)a^{\dagger}\sigma_{2}^{+}+g_{2}\cosh\left(r\right)a\sigma_{2}^{+}+\text{h.c.},\label{eq:Hamiltonian_2}
\end{align}
where $g_{1}\equiv\sqrt{|g_{1,\text{b}}|^{2}-|g_{1,\text{r}}|^{2}}$
and $g_{2}\equiv\sqrt{|g_{2,\text{r}}|^{2}-|g_{2,\text{b}}|^{2}}$
denote effective coupling strengths. The physical significance of
this reparameterization becomes evident when applying the squeezing
transformation to the mode operators. Specifically, the system is
formally equivalent to a standard laser model $H_{sq}=g_{1}b^{\dag}\sigma_{1}^{+}+g_{2}b\sigma_{2}^{+}+\text{h.c.}$
acting on the squeezed mode defined by $b=SaS^{\dag}.$

In this representation, $H_{sq}$ takes the exact form of the dual-species
Hamiltonian implemented by Ref.~\citep{Behrle2023}, where gain and
loss are independently provided by heating (AJC) and cooling (JC)
interactions, respectively. While the setup in Ref.~\citep{Behrle2023}
operates on the bare motional mode, our model demonstrates that by
leveraging the intrinsic ion-phonon interactions via dual-tone driving,
the same laser-like dynamics are engineered to act directly on a squeezed
mode. This approach allows for the generation of squeezed lasing in
the laboratory basis without the need for additional external reservoir
engineering.

In the following sections, we numerically explore the steady-state
behavior of the model across a range of parameters. We first analyze
the onset of lasing and the associated nonclassical phonon statistics,
including gain-loss balance, intensity correlations, and spectral
narrowing. We then examine the role of the squeezing parameter in
shaping the statistical properties of the phonon field. Finally, we
demonstrate how an external coherent drive can break the intrinsic
phase symmetry and enable controlled quadrature squeezing of the phonon
mode.

\section{Signatures of Squeezed Lasing}

\begin{figure}[h]
\begin{centering}
\includegraphics[width=8.8cm]{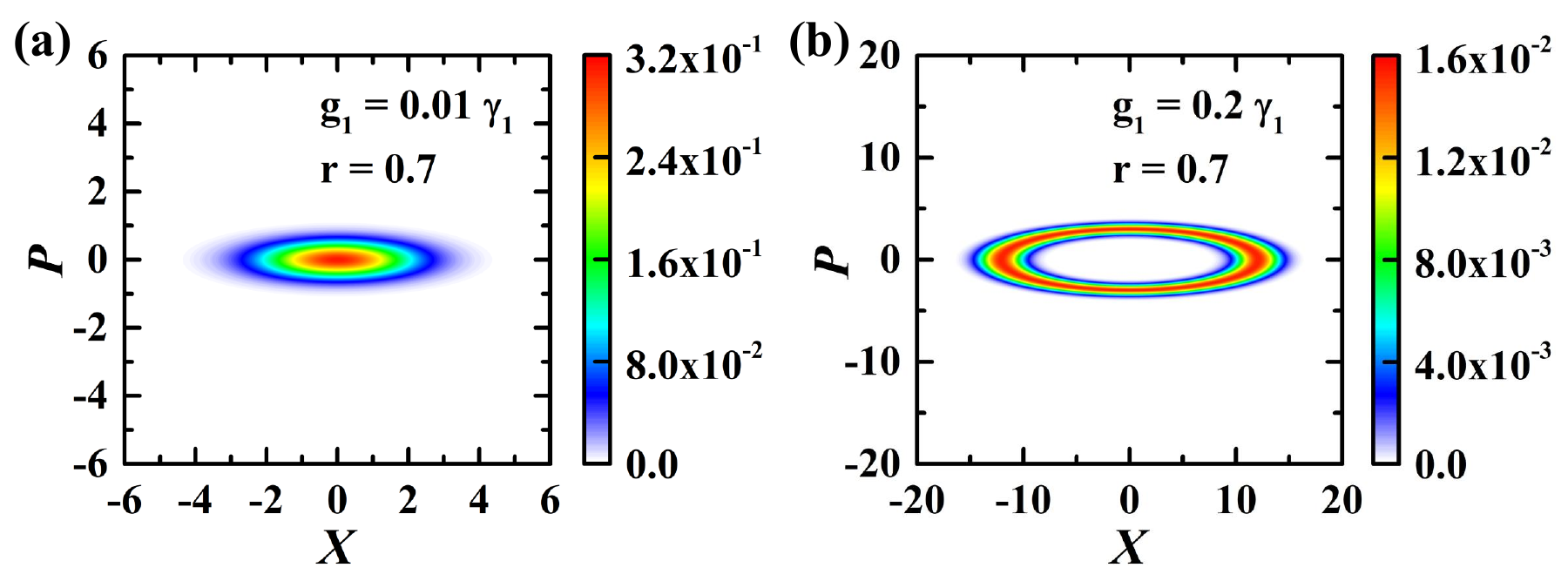}
\par\end{centering}
\caption{Steady-state Wigner functions of the phonon mode. (a) Below-threshold
regime at $g_{1}=0.01\gamma_{1},$ exhibiting a near-symmetric profile
characteristic of a squeezed vacuum state. (b) Above-threshold regime
at $g_{1}=0.2\gamma_{1},$ showing a displaced, elliptically deformed
structure consistent with a random-phase squeezed coherent state.
Simulation parameters are fixed at $g_{2}=0.15\gamma_{1}$ and $\gamma_{2}=2.5\gamma_{1}.$
\label{fig:wigner}}
\end{figure}

The transition into the squeezed-lasing regime is characterized by
the emergence of a displaced squeezed coherent state from an initial
squeezed vacuum. We begin our analysis by examining the system's behavior
in the limit $g_{1}\rightarrow0.$ In this regime, as reproduced in
Fig.~\ref{fig:wigner}(a), the model reduces to an incoherent phonon-pumping
scheme previously studied in the context of dark squeezed states \citep{Cirac1993}.
However, as the effective heating coupling $g_{1}$ increases and
the system enters the lasing regime, the steady-state Wigner function
undergoes a fundamental transformation. As shown in Fig.~\ref{fig:wigner}(b),
the distribution develops a displaced, elliptically deformed structure,
corresponding to a squeezed coherent state with a randomly selected
phase.

To quantify the lasing transition, we first analyze the effective
gain and loss mechanisms induced by the driven ions. Following the
adiabatic elimination of the internal dynamics (Appendix~\ref{sec:gain}),
the gain and decay operators are given by

\begin{align}
\mathcal{G} & =-\frac{\left|g_{1}\right|^{2}}{\gamma_{1}}\sigma_{1,z}, & \mathcal{K}=-\frac{\left|g_{2}\right|^{2}}{\gamma_{2}}\sigma_{2,z},\label{eq:gain_decay}
\end{align}
where $\gamma_{1}$ and $\gamma_{2}$ are the spontaneous decay rates
of the two ions. A crucial physical insight of our model is that these
operators depend solely on the effective coupling strengths and internal
decay rates, remaining entirely independent of the squeezing parameter
$r$. This independence confirms that while squeezing reshapes the
mode structure, it does not directly alter the net gain-loss balance.

\begin{figure}[h]
\begin{centering}
\includegraphics[width=8.8cm]{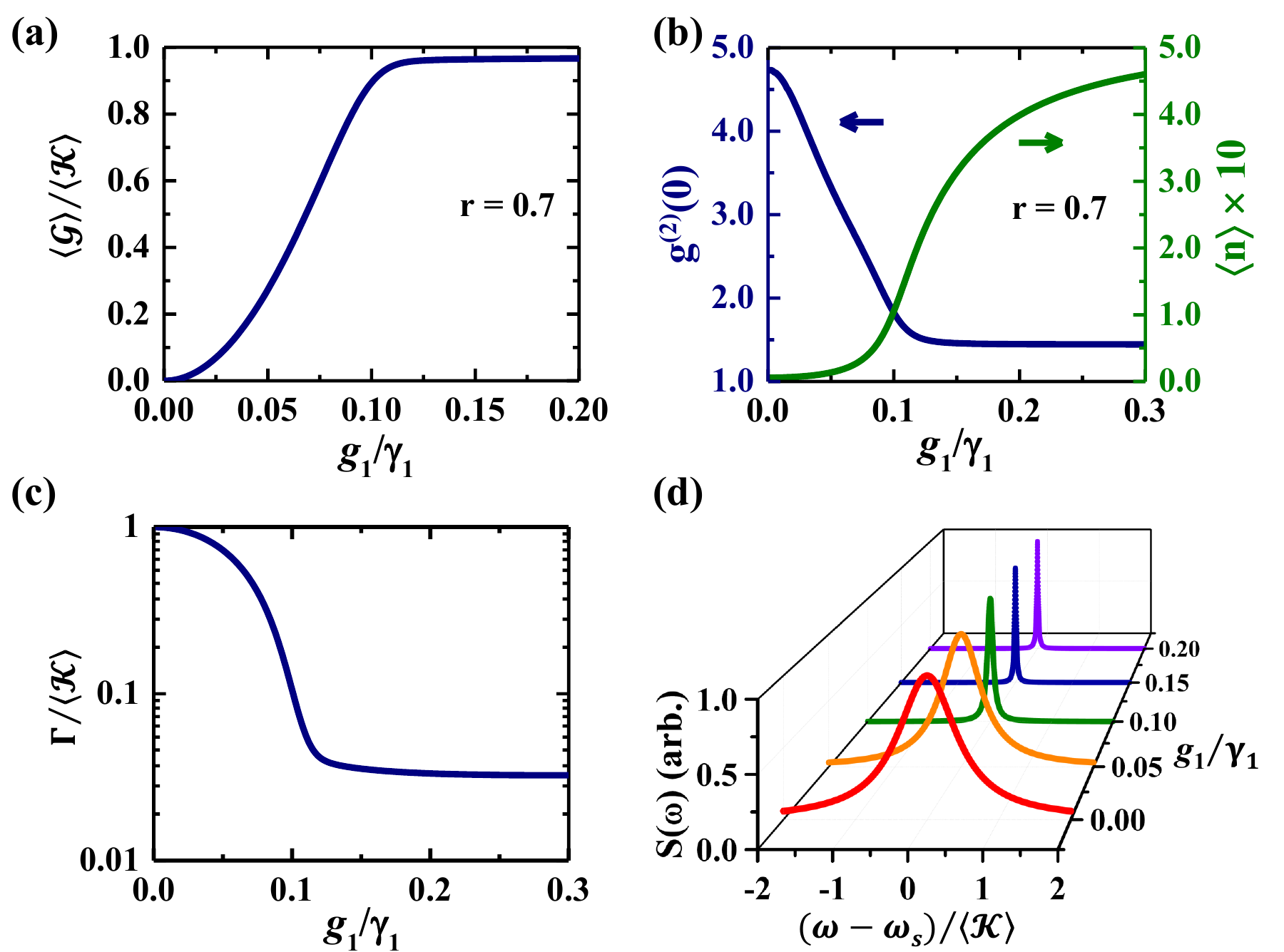}
\par\end{centering}
\caption{Signatures of the squeezed-lasing transition. (a) Steady-state ratio
of gain to loss, $\left\langle \mathcal{G}\right\rangle /\left\langle \mathcal{K}\right\rangle ,$
as a function of the effective heating coupling $g_{1},$ with the
threshold identified at $g_{1,\text{th}}\approx0.12\gamma_{1}.$ (b)
Second-order coherence $g^{(2)}\left(0\right)$ (left axis) and mean
phonon number $\left\langle n\right\rangle $ (right axis) of the
phonon mode as functions of $g_{1}.$ (c) Emission linewidth $\Gamma$
verse $g_{1},$ demonstrating linewidth narrowing above the threshold.
(d) Normalized emission spectrum $S\left(\omega\right)$ for various
$\nicefrac{g_{1}}{\gamma_{1}},$ where frequency is defined relative
to the squeezed mode frequency $\omega_{s}.$ \label{fig:laser_property}}
\end{figure}

In Fig.~\ref{fig:laser_property}(a), we plot the steady-state ratio
$\left\langle \mathcal{G}\right\rangle /\left\langle \mathcal{K}\right\rangle $
as a function of the effective heating coupling $g_{1}.$ A distinct
threshold behavior is observed near the critical point $g_{1,\text{th}}\approx0.12\gamma_{1},$
above which the gain overcomes the losses and the system transitions
into the lasing regime.

The establishment of coherence is further evidenced by the steady-state
second-order correlation function $g^{\left(2\right)}\left(0\right)\equiv\left.\nicefrac{\left\langle a^{\dag}\left(0\right)a^{\dag}\left(\tau\right)a\left(\tau\right)a\left(0\right)\right\rangle }{\left\langle a^{\dag}\left(\tau\right)a\left(\tau\right)\right\rangle ^{2}}\right|_{\tau=0}$
and the mean phonon number $\left\langle n\right\rangle =\left\langle a^{\dag}a\right\rangle ,$
shown in Fig.~\ref{fig:laser_property}(b). As $g_{1}$ surpasses
the threshold, $g^{\left(2\right)}\left(0\right)$ decreases while
$\left\langle n\right\rangle $ rises, indicating the emergence of
coherence from a nonclassical vacuum state. This transition is accompanied
by linewidth narrowing in the emission spectrum, which follows a Lorentzian
profile:

\begin{equation}
S\left(\omega\right)=\frac{n_{\text{ss}}}{\omega^{2}+\Gamma^{2}/4}\label{eq:spectrum}
\end{equation}
where $n_{\text{ss}}$ is the steady-state phonon number and $\Gamma=\left\langle \mathcal{K}\right\rangle -\left\langle \mathcal{G}\right\rangle $
represents the effective emission linewidth. Figures~\ref{fig:laser_property}(c)
and (d) clearly illustrate linewidth narrowing above the lasing threshold,
confirming the buildup of phase coherence as the gain approaches the
loss rate.

\begin{figure}[h]
\begin{centering}
\includegraphics[width=8.8cm]{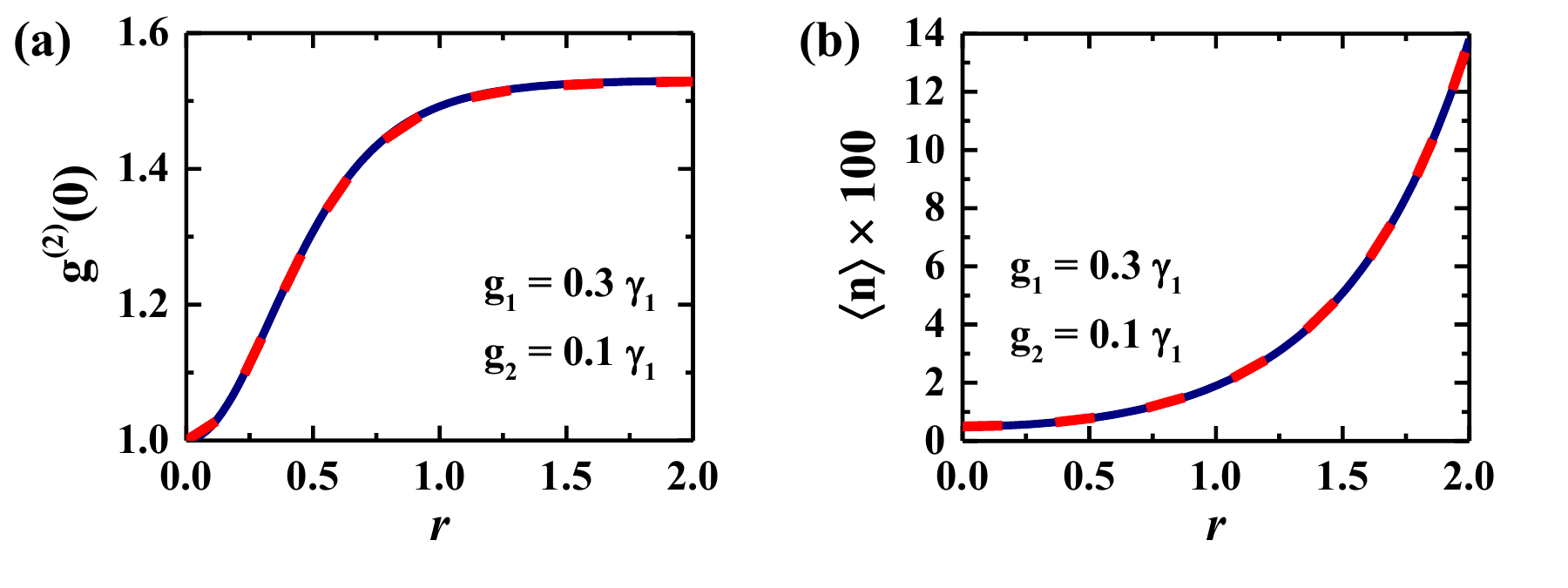}
\par\end{centering}
\caption{Dependence of phonon statistics on the squeezing parameter $r$ in
the above-threshold regime. (a) Second-order coherence function $g^{(2)}(0)$
and (b) mean phonon number $\left\langle n\right\rangle $ as functions
of $r.$ Solid curves represent numerical results from the master
equation, while dashed curves indicate analytical predictions based
on the mapping to the squeezed frame. All results are obtained well
above the lasing threshold. \label{fig:squeezing_para}}
\end{figure}

A key advantage of our \textquotedbl bath-free\textquotedbl{} scheme
is that the squeezing parameter $r$ can be continuously tuned via
the ratio of red and blue sideband couplings. To investigate its impact
on lasing statistics, we fix the system well above threshold and explore
the dependence of $\left\langle n\right\rangle $ and $g^{(2)}\left(0\right)$
on $r$ in Fig.~\ref{fig:squeezing_para}. The analytical expressions
for these statistical properties are given by
\begin{align}
\left\langle n\right\rangle  & =\left|\alpha_{s}\right|^{2}\cosh\left(2r\right)+\sinh^{2}\left(r\right),\label{eq:n}\\
g^{(2)}\left(0\right) & =\frac{1}{4\left\langle n\right\rangle ^{2}}\left[3\left|\alpha_{s}\right|^{2}\left(\left|\alpha_{s}\right|^{2}+2\right)\cosh\left(4r\right)\right.\nonumber \\
 & \left.+\left(3\cosh\left(2r\right)-8\left|\alpha_{s}\right|^{2}-4\right)\cosh\left(2r\right)+\left(\left|\alpha_{s}\right|^{2}+1\right)^{2}\right],\label{eq:g2}
\end{align}
where $\left|\alpha_{s}\right|^{2}$ denotes the coherent excitation
amplitude in the squeezed frame. As shown in Fig.~\ref{fig:squeezing_para},
our numerical results show excellent agreement with these analytical
predictions. In the strongly squeezed regime $\left(r\gg1\right),$
$g^{(2)}\left(0\right)$ saturates to approximately 1.5, reflecting
an intermediate phonon bunching behavior that exceeds the fluctuations
of a standard coherent state but remains below those of a thermal
state.

\section{Phase Symmetry Breaking and Quadrature Squeezing\label{sec:Phase-Symmetry-Breaking}}

While the squeezed-lasing regime inherently exhibits a randomly distributed
phase, the phase coherence of the emitted phonon field can be stabilized
by applying an external drive with a well-defined reference. Experimentally,
this is implemented by applying two resonant oscillating forces directly
to the phonon mode via an electrode of the ion trap. The resulting
driving Hamiltonian takes the form $H_{\text{d,s}}=g_{\text{d,}1}\left(ae^{i\phi_{1}}+a^{\dag}e^{-i\phi_{1}}\right)+g_{\text{d,}2}\left(ae^{i\phi_{2}}+a^{\dag}e^{-i\phi_{2}}\right),$
where $g_{\text{d,}1},g_{\text{d,}2}$ and $\phi_{1},\phi_{2}$ denote
the amplitudes and phases of the two coherent drives, respectively.
By appropriately choosing the amplitude ratio and relative phase of
the driving forces, the external drive can be engineered to match
the same squeezing parameters $r$ and $\theta$ that define the internal
squeezed-lasing dynamics. Specifically, we identify the matching conditions
as $\tanh\left(r\right)=\left|\frac{g_{\text{d},2}}{g_{\text{d},1}}\right|<1,$
and $\theta=-\phi_{1}-\phi_{2}.$

To avoid overwhelming the intrinsic lasing behavior and to ensure
compatibility with the squeezed-lasing steady state, we choose weak
driving amplitudes, such as $g_{\text{d},1}=0.015\gamma_{1}$ and
$g_{\text{d},2}=0.009\gamma_{1},$ corresponding to a moderate squeezing
amplitude $r=0.7.$

\begin{figure}[h]
\begin{centering}
\includegraphics[width=8.8cm]{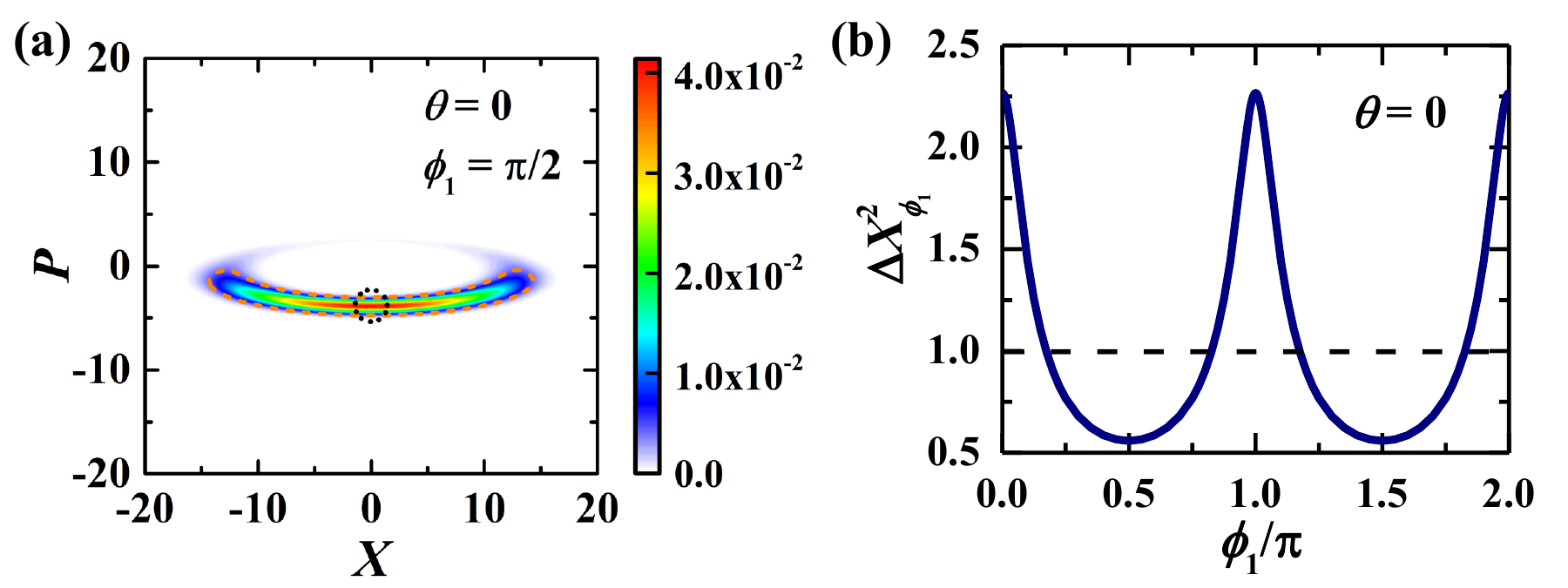}
\par\end{centering}
\caption{Phase-symmetry breaking and quadrature squeezing. (a) Steady-state
Wigner function in the presence of a weak phase-stabilizing field.
Dashed orange lines indicate the 0.1 contour of the Wigner distribution,
compared to the black dotted contours of a coherent state with equal
amplitude. (b) Quadrature variance $\Delta X_{\phi_{1}}^{2}$ as a
function of the drive phase $\phi_{1},$ with the dashed line marking
the vacuum-level fluctuations. Quadrature squeezing $\left(\Delta X_{\phi_{1}}^{2}<1\right)$
is observed near $\phi_{1}=0.5\pi$ and $1.5\pi.$ Shared parameters:
$g_{1}=0.2\gamma_{1},g_{2}=0.15\gamma_{1},\gamma_{2}=2.5\gamma_{1},\theta=0;$
panel (a) uses $\phi_{1}=\pi/2.$ \label{fig:phase_locked}}
\end{figure}

The influence of the external drive is clearly visible in the steady-state
Wigner function shown in Fig.~\ref{fig:phase_locked}(a). The application
of these weak phase-stabilizing forces results in a displacement of
the phonon field in phase space and breaks the continuous phase symmetry,
leading to a well-defined phase orientation for the squeezed state.

To probe phase-sensitive properties, we vary the drive phase $\phi_{1}$
while keeping the system well above the lasing threshold. The corresponding
quadrature fluctuations, measured along different phase axes, exhibit
a strong dependence on $\phi_{1}.$ We define the generalized quadrature
operator as $X_{\phi_{1}}\equiv ae^{-i\phi_{1}}+a^{\dag}e^{i\phi_{1}},$
where the standard quadratures correspond to $X=X_{0}$ and $P=X_{\pi/2}.$
For reference, the vacuum fluctuation level is normalized to $\Delta X_{\phi_{1}}^{2}=1.$

As shown in Fig.~\ref{fig:phase_locked}(b), the quadrature variance
drops below the vacuum level for specific phase values near $\phi_{\text{d},1}=0.5\pi,$
and $1.5\pi,$ indicating the presence of genuine quadrature squeezing.
This demonstrates that the external drive not only stabilizes the
phase but also enables precise control over the squeezing axis. Such
control allows noise suppression to be optimized along targeted quadrature
directions, a feature particularly valuable for practical applications
in quantum sensing.

\section{Experimental Implementation in a Trapped-Ion System}

Our theoretical model assumes two effective two-level systems coupled
to a shared vibrational mode and driven on both red- and blue-sideband
transitions {[}see Eq.~\ref{eq:Hamiltonain}{]}. We now outline how
this situation can be engineered in existing trapped-ion platforms,
and in particular how it can be naturally implemented by modifying
the two-species phonon-laser setup of Behrle \textit{et al.}~\citep{Behrle2023}.

In that experiment, a mixed chain of $^{40}\text{Ca}^{+}$ and $^{9}\text{Be}^{+}$
ions is confined in a linear Paul trap and coupled to a common axial
motional mode with frequency $\omega_{m}$ in the megahertz range.
The calcium ion realizes a JC interaction on the red sideband of an
optical qubit transition, while the beryllium ion realizes an AJC
interaction on the blue sideband of a hyperfine qubit. Optical pumping
via short-lived excited states provides controlled dissipation for
each ion, leading to competing cooling and heating channels for the
shared phonon mode and giving rise to a phonon-lasing transition at
low mean phonon numbers $\bar{n}\lesssim10.$

To realize the squeezed-lasing Hamiltonian in Eq.~\ref{eq:Hamiltonian_2},
each ion must experience both red- and blue-sideband couplings with
independently tunable amplitudes $g_{i,\text{r}}$ and $g_{i,\text{b}}.$
This can be implemented by adding an extra sideband tone to each ion
in the above configuration. For the \textquotedblleft cooling\textquotedblright{}
ion, one supplements the red-sideband laser with a weaker blue-sideband
tone at the same trap frequency, realizing effective couplings $\left(g_{2,\text{r}},g_{2,\text{b}}\right).$
Likewise, for the \textquotedblleft heating\textquotedblright{} ion
one adds a red-sideband tone in addition to the existing blue-sideband
drive, realizing $\left(g_{1,\text{r}},g_{1,\text{b}}\right).$ By
adjusting the relative intensities and optical phases, the squeezing
parameter and phase, $\tanh\left(r\right)=\left|\frac{g_{1,\text{r}}}{g_{1,\text{b}}}\right|=\left|\frac{g_{2,\text{b}}}{g_{2,\text{r}}}\right|<1,$
and $\theta=\phi_{1,\text{r}}-\phi_{1,\text{b}}=\phi_{2,\text{r}}-\phi_{2,\text{b}}$
can be tuned to the desired values. 

The dissipative ingredients required for our gain and loss operators
$\mathcal{G}$ and $\mathcal{K}$ in Eq.~\ref{eq:gain_decay} are
already present: optical pumping on the calcium and beryllium ions
sets the decay rates $\gamma_{2}$ and $\gamma_{1},$ respectively.
In the regime $\gamma_{1,2}\gg g_{1,2},$ which is satisfied for sideband
Rabi frequencies of a few kilohertz and optical pumping rates in the
tens to hundreds of kilohertz, the internal dynamics can be adiabatically
eliminated and the phonon mode obeys the simple gain\textendash loss
master equation used in our analysis. Typical trap parameters (axial
frequencies in the megahertz range and Lamb\textendash Dicke parameters
of order $10^{-2}-10^{-1})$ permit tuning of the sideband couplings
over at least an order of magnitude via the laser intensities. This
allows one to access both the below-threshold regime, where the phonon
mode is close to a squeezed-vacuum\textendash like state, and the
above-threshold regime, where our theory predicts a displaced squeezed
steady state with partial phonon bunching and linewidth narrowing.

Phase locking of the lasing mode can be realized by applying a weak
resonant oscillating force to a trap electrode, providing the coherent
drive used in Sec.~\ref{sec:Phase-Symmetry-Breaking}. Matching the
drive phase to the intrinsic squeezing parameters $r$ and $\theta$
implements the symmetry-breaking mechanism that stabilizes a particular
quadrature of motion and allows direct measurements of quadrature
squeezing. Importantly, all of these steps involve only coherent sideband
drives and standard optical pumping\textemdash no squeezed reservoirs,
parametric cavity drives, or other forms of reservoir engineering
are required to realize squeezed phonon lasing in this setup.

\section{Discussion and Conclusion}

Our results demonstrate that squeezed-lasing behavior can be achieved
in a minimal trapped-ion system consisting of two ions coupled to
a shared vibrational mode, without the need for engineered dissipation
or external reservoirs. This approach contrasts with previous schemes
for squeezed lasers in optical and superconducting systems, which
typically rely on engineered baths to shape the steady-state quantum
statistics of the emitted field. In our approach, squeezed lasing
emerges purely from coherent ion-phonon interactions, achieved by
simultaneously driving the ions on their red- and blue-sideband transitions.
These interactions induce effective squeezed couplings between the
ions and the phonon mode, generating the desired squeezed state through
intrinsic control mechanisms.

By analyzing the steady-state behavior of the system under varying
coupling strengths, we identify a clear lasing threshold characterized
by the balance of gain and loss, the buildup of coherent excitation,
and spectral narrowing. Above threshold, the system supports a squeezed
coherent phonon state with tunable statistical properties. The squeezing
parameter $r$ can be adjusted by varying the ratio of red- and blue-sideband
couplings, providing direct control over both the mean phonon number
and the second-order correlation function. In the strongly squeezed
regime, we observe partial phonon bunching, which is indicative of
a nonclassical, non-Poissonian steady state.

Furthermore, we have shown that the intrinsic phase symmetry of the
system can be explicitly broken by applying a weak external drive.
This drive not only selects a well-defined phase for the lasing mode
but also facilitates the observation of quadrature squeezing below
the vacuum level. This feature is particularly valuable for practical
applications, as it enables noise suppression in targeted quadratures
while maintaining overall coherence.

Taken together, these findings establish a new direction for realizing
nonclassical light-like states in trapped-ion systems. By eliminating
the need for bath engineering, our work simplifies both the theoretical
modeling and experimental implementation, offering a cleaner and more
controllable approach to exploring squeezed-lasing physics. Our approach
opens new possibilities for integrating squeezed phonon states into
quantum sensing and information protocols, and extends the concept
of squeezed lasing to other isolated bosonic systems beyond optics.

\section*{Acknowledgments}

We acknowledge supportfromthe National Science and Technology Council
(NSTC), Taiwan, under Grant No. NSTC-113-2112-M-002-025 and No. NSTC-112-2112-M-002-001.
We are also grateful for support from TG 1.2 of NCTS at NTU.

\appendix
\renewcommand{\appendixname}{APPENDIX}

\section{Derivation of Gain, Decay, and Linewidth\label{sec:gain}}

In this appendix, we derive the gain, decay, and linewidth expressions
used to characterize the spectral properties of the squeezed-lasing
system.

\subsection{Gain and Decay Operators}

We begin by examining the equations of motion for the system operators
in the Heisenberg picture, derived from the Hamiltonian in Eq.~\ref{eq:Hamiltonian_2}.
The evolution of the phonon annihilation operator a and the internal
atomic operators is given by:
\begin{align*}
\partial_{t}a & =-ig_{1}\cosh\left(r\right)\sigma_{1}^{+}-ig_{1}^{*}\sinh\left(r\right)\sigma_{1}^{-}\\
 & -ig_{2}\sinh\left(r\right)\sigma_{2}^{+}-ig_{2}^{*}\cosh\left(r\right)\sigma_{2}^{-},\\
\partial_{t}\sigma_{1}^{-} & =-\frac{\gamma_{1}}{2}\sigma_{1}^{-}+ig_{1}\cosh\left(r\right)a^{\dag}\sigma_{1,z}+ig_{1}\sinh\left(r\right)\sigma_{1,z}a,\\
\partial_{t}\sigma_{2}^{-} & =-\frac{\gamma_{2}}{2}\sigma_{2}^{-}+ig_{2}\sinh\left(r\right)a^{\dag}\sigma_{2,z}+ig_{2}\cosh\left(r\right)\sigma_{2,z}a,\\
\partial_{t}\sigma_{1,z} & =-\gamma_{1}\left(\sigma_{1,z}+1\right)-2ig_{1}\cosh\left(r\right)a^{\dag}\sigma_{1}^{+}+2ig_{1}^{*}\cosh\left(r\right)a\sigma_{1}^{-}\\
 & -2ig_{1}\sinh\left(r\right)a\sigma_{1}^{+}+2ig_{1}^{*}\sinh\left(r\right)a^{\dagger}\sigma_{1}^{-},\\
\partial_{t}\sigma_{2,z} & =-\gamma_{2}\left(\sigma_{2,z}+1\right)-2ig_{2}\sinh\left(r\right)a^{\dag}\sigma_{2}^{+}+2ig_{2}^{*}\sinh\left(r\right)a\sigma_{2}^{-}\\
 & -2ig_{2}\cosh\left(r\right)a\sigma_{2}^{+}+2ig_{2}^{*}\cosh\left(r\right)a^{\dagger}\sigma_{2}^{-}.
\end{align*}
We consider the regime $\gamma_{1,2}\gg g_{1,2},$ where the internal
dynamics of the ions reach a steady state on a time scale much shorter
than the evolution of the phonon mode. Therefore, we can assume the
steady state for the spin variables ($\partial_{t}\sigma_{1/2}^{-}=0$
and $\partial_{t}\sigma_{1/2,z}=0$) and obtain the following evolution
equation for the phonon mode:
\begin{align*}
\partial_{t}a & =-\frac{\left|g_{1}\right|^{2}}{\frac{\gamma_{1}}{2}}\sigma_{1,z}a+\frac{\left|g_{2}\right|^{2}}{\frac{\gamma_{2}}{2}}\sigma_{2,z}a\\
 & \equiv\frac{1}{2}\mathcal{G}a-\frac{1}{2}\mathcal{K}a,
\end{align*}
where $\mathcal{G}$ and $\mathcal{K}$ are the gain and decay operators,
respectively. They are defined as
\begin{align*}
\mathcal{G} & =-\frac{\left|g_{1}\right|^{2}}{\gamma_{1}}\sigma_{1,z},\\
\mathcal{K} & =-\frac{\left|g_{2}\right|^{2}}{\gamma_{2}}\sigma_{2,z}.
\end{align*}
These operators represent the effective gain and decay rates due to
the interaction of the ions with the phonon mode.

\subsection{Laser Linewidth}

To calculate the laser linewidth, we can perform the Fourier transformation
of the two-time correlation function $\left\langle a^{\dagger}\left(\tau\right)a\left(0\right)\right\rangle .$
Since the system is in the steady state, the expectation values of
both gain and decay operators are constant, and the equation for $a\left(t\right)$
is given by:
\[
\dot{a}\left(t\right)=\frac{1}{2}\left(\left\langle \mathcal{G}\right\rangle -\left\langle \mathcal{K}\right\rangle \right)a\left(t\right).
\]
The corresponding correlation function is:
\begin{align*}
\left\langle a^{\dagger}\left(\tau\right)a\left(0\right)\right\rangle  & =\left\langle a^{\dagger}\left(0\right)a\left(0\right)\right\rangle e^{\frac{1}{2}\left(\left\langle \mathcal{G}\right\rangle -\left\langle \mathcal{K}\right\rangle \right)\tau}\\
 & \equiv\left\langle a^{\dagger}\left(0\right)a\left(0\right)\right\rangle e^{-\frac{\Gamma}{2}\tau},
\end{align*}
where the linewidth is $\Gamma=\left(\left\langle \mathcal{K}\right\rangle -\left\langle \mathcal{G}\right\rangle \right).$
The spectral lineshape of the system is then given by:
\[
S\left(\omega\right)=\left\langle a^{\dagger}\left(0\right)a\left(0\right)\right\rangle \mathcal{F}\left(e^{-\frac{\Gamma}{2}\tau}\right)=\frac{n_{\text{ss}}}{\omega^{2}+\frac{\Gamma^{2}}{4}},
\]
where $n_{\text{ss}}$ is the steady-state phonon number and $\mathcal{F}$
is the Fourier transform operation.

\bibliographystyle{apsrev4-2}
\bibliography{ms}

\begin{thebibliography}{41}%
\makeatletter
\providecommand \@ifxundefined [1]{%
 \@ifx{#1\undefined}
}%
\providecommand \@ifnum [1]{%
 \ifnum #1\expandafter \@firstoftwo
 \else \expandafter \@secondoftwo
 \fi
}%
\providecommand \@ifx [1]{%
 \ifx #1\expandafter \@firstoftwo
 \else \expandafter \@secondoftwo
 \fi
}%
\providecommand \natexlab [1]{#1}%
\providecommand \enquote  [1]{``#1''}%
\providecommand \bibnamefont  [1]{#1}%
\providecommand \bibfnamefont [1]{#1}%
\providecommand \citenamefont [1]{#1}%
\providecommand \href@noop [0]{\@secondoftwo}%
\providecommand \href [0]{\begingroup \@sanitize@url \@href}%
\providecommand \@href[1]{\@@startlink{#1}\@@href}%
\providecommand \@@href[1]{\endgroup#1\@@endlink}%
\providecommand \@sanitize@url [0]{\catcode `\\12\catcode `\$12\catcode `\&12\catcode `\#12\catcode `\^12\catcode `\_12\catcode `\%12\relax}%
\providecommand \@@startlink[1]{}%
\providecommand \@@endlink[0]{}%
\providecommand \url  [0]{\begingroup\@sanitize@url \@url }%
\providecommand \@url [1]{\endgroup\@href {#1}{\urlprefix }}%
\providecommand \urlprefix  [0]{URL }%
\providecommand \Eprint [0]{\href }%
\providecommand \doibase [0]{https://doi.org/}%
\providecommand \selectlanguage [0]{\@gobble}%
\providecommand \bibinfo  [0]{\@secondoftwo}%
\providecommand \bibfield  [0]{\@secondoftwo}%
\providecommand \translation [1]{[#1]}%
\providecommand \BibitemOpen [0]{}%
\providecommand \bibitemStop [0]{}%
\providecommand \bibitemNoStop [0]{.\EOS\space}%
\providecommand \EOS [0]{\spacefactor3000\relax}%
\providecommand \BibitemShut  [1]{\csname bibitem#1\endcsname}%
\let\auto@bib@innerbib\@empty
\bibitem [{\citenamefont {Cirac}\ and\ \citenamefont {Zoller}(1995)}]{Cirac1995}%
  \BibitemOpen
  \bibfield  {author} {\bibinfo {author} {\bibfnamefont {J.~I.}\ \bibnamefont {Cirac}}\ and\ \bibinfo {author} {\bibfnamefont {P.}~\bibnamefont {Zoller}},\ }\href {https://doi.org/10.1103/PhysRevLett.74.4091} {\bibfield  {journal} {\bibinfo  {journal} {Phys. Rev. Lett.}\ }\textbf {\bibinfo {volume} {74}},\ \bibinfo {pages} {4091} (\bibinfo {year} {1995})}\BibitemShut {NoStop}%
\bibitem [{\citenamefont {Soderberg}\ and\ \citenamefont {Monroe}(2010)}]{Soderberg2010}%
  \BibitemOpen
  \bibfield  {author} {\bibinfo {author} {\bibfnamefont {K.-A.~B.}\ \bibnamefont {Soderberg}}\ and\ \bibinfo {author} {\bibfnamefont {C.}~\bibnamefont {Monroe}},\ }\href {https://doi.org/10.1088/0034-4885/73/3/036401} {\bibfield  {journal} {\bibinfo  {journal} {Rep. Prog. Phys.}\ }\textbf {\bibinfo {volume} {73}},\ \bibinfo {pages} {036401} (\bibinfo {year} {2010})}\BibitemShut {NoStop}%
\bibitem [{\citenamefont {Ruskov}\ and\ \citenamefont {Tahan}(2012)}]{Ruskov2012}%
  \BibitemOpen
  \bibfield  {author} {\bibinfo {author} {\bibfnamefont {R.}~\bibnamefont {Ruskov}}\ and\ \bibinfo {author} {\bibfnamefont {C.}~\bibnamefont {Tahan}},\ }\href {https://doi.org/10.1088/1742-6596/398/1/012011} {\bibfield  {journal} {\bibinfo  {journal} {J. Phys: Conf. Ser.}\ }\textbf {\bibinfo {volume} {398}},\ \bibinfo {pages} {012011} (\bibinfo {year} {2012})}\BibitemShut {NoStop}%
\bibitem [{\citenamefont {Reinke}\ and\ \citenamefont {El-Kady}(2016)}]{Reinke2016}%
  \BibitemOpen
  \bibfield  {author} {\bibinfo {author} {\bibfnamefont {C.~M.}\ \bibnamefont {Reinke}}\ and\ \bibinfo {author} {\bibfnamefont {I.}~\bibnamefont {El-Kady}},\ }\href {https://doi.org/10.1063/1.4972568} {\bibfield  {journal} {\bibinfo  {journal} {AIP Adv.}\ }\textbf {\bibinfo {volume} {6}},\ \bibinfo {pages} {122002} (\bibinfo {year} {2016})}\BibitemShut {NoStop}%
\bibitem [{\citenamefont {Fl{\"u}hmann}\ \emph {et~al.}(2019)\citenamefont {Fl{\"u}hmann}, \citenamefont {Nguyen}, \citenamefont {Marinelli}, \citenamefont {Negnevitsky}, \citenamefont {Mehta},\ and\ \citenamefont {Home}}]{Fluehmann2019}%
  \BibitemOpen
  \bibfield  {author} {\bibinfo {author} {\bibfnamefont {C.}~\bibnamefont {Fl{\"u}hmann}}, \bibinfo {author} {\bibfnamefont {T.~L.}\ \bibnamefont {Nguyen}}, \bibinfo {author} {\bibfnamefont {M.}~\bibnamefont {Marinelli}}, \bibinfo {author} {\bibfnamefont {V.}~\bibnamefont {Negnevitsky}}, \bibinfo {author} {\bibfnamefont {K.}~\bibnamefont {Mehta}},\ and\ \bibinfo {author} {\bibfnamefont {J.~P.}\ \bibnamefont {Home}},\ }\href {https://doi.org/10.1038/s41586-019-0960-6} {\bibfield  {journal} {\bibinfo  {journal} {Nature}\ }\textbf {\bibinfo {volume} {566}},\ \bibinfo {pages} {513} (\bibinfo {year} {2019})}\BibitemShut {NoStop}%
\bibitem [{\citenamefont {Ge}\ \emph {et~al.}(2019)\citenamefont {Ge}, \citenamefont {Sawyer}, \citenamefont {Britton}, \citenamefont {Jacobs}, \citenamefont {Bollinger},\ and\ \citenamefont {Foss-Feig}}]{Ge2019}%
  \BibitemOpen
  \bibfield  {author} {\bibinfo {author} {\bibfnamefont {W.}~\bibnamefont {Ge}}, \bibinfo {author} {\bibfnamefont {B.~C.}\ \bibnamefont {Sawyer}}, \bibinfo {author} {\bibfnamefont {J.~W.}\ \bibnamefont {Britton}}, \bibinfo {author} {\bibfnamefont {K.}~\bibnamefont {Jacobs}}, \bibinfo {author} {\bibfnamefont {J.~J.}\ \bibnamefont {Bollinger}},\ and\ \bibinfo {author} {\bibfnamefont {M.}~\bibnamefont {Foss-Feig}},\ }\href {https://doi.org/10.1103/physrevlett.122.030501} {\bibfield  {journal} {\bibinfo  {journal} {Phys. Rev. Lett.}\ }\textbf {\bibinfo {volume} {122}},\ \bibinfo {pages} {030501} (\bibinfo {year} {2019})}\BibitemShut {NoStop}%
\bibitem [{\citenamefont {Loye}\ \emph {et~al.}(2020)\citenamefont {Loye}, \citenamefont {Lages},\ and\ \citenamefont {Shepelyansky}}]{Loye2020}%
  \BibitemOpen
  \bibfield  {author} {\bibinfo {author} {\bibfnamefont {J.}~\bibnamefont {Loye}}, \bibinfo {author} {\bibfnamefont {J.}~\bibnamefont {Lages}},\ and\ \bibinfo {author} {\bibfnamefont {D.~L.}\ \bibnamefont {Shepelyansky}},\ }\href {https://doi.org/10.1103/physreva.101.032349} {\bibfield  {journal} {\bibinfo  {journal} {Phys. Rev. A}\ }\textbf {\bibinfo {volume} {101}},\ \bibinfo {pages} {032349} (\bibinfo {year} {2020})}\BibitemShut {NoStop}%
\bibitem [{\citenamefont {Gan}\ \emph {et~al.}(2020)\citenamefont {Gan}, \citenamefont {Maslennikov}, \citenamefont {Tseng}, \citenamefont {Nguyen},\ and\ \citenamefont {Matsukevich}}]{Gan2020}%
  \BibitemOpen
  \bibfield  {author} {\bibinfo {author} {\bibfnamefont {H.}~\bibnamefont {Gan}}, \bibinfo {author} {\bibfnamefont {G.}~\bibnamefont {Maslennikov}}, \bibinfo {author} {\bibfnamefont {K.-W.}\ \bibnamefont {Tseng}}, \bibinfo {author} {\bibfnamefont {C.}~\bibnamefont {Nguyen}},\ and\ \bibinfo {author} {\bibfnamefont {D.}~\bibnamefont {Matsukevich}},\ }\href {https://doi.org/10.1103/physrevlett.124.170502} {\bibfield  {journal} {\bibinfo  {journal} {Physical Review Letters}\ }\textbf {\bibinfo {volume} {124}},\ \bibinfo {pages} {170502} (\bibinfo {year} {2020})}\BibitemShut {NoStop}%
\bibitem [{\citenamefont {Nguyen}\ \emph {et~al.}(2021)\citenamefont {Nguyen}, \citenamefont {Tseng}, \citenamefont {Maslennikov}, \citenamefont {Gan},\ and\ \citenamefont {Matsukevich}}]{Nguyen2021}%
  \BibitemOpen
  \bibfield  {author} {\bibinfo {author} {\bibfnamefont {C.-H.}\ \bibnamefont {Nguyen}}, \bibinfo {author} {\bibfnamefont {K.-W.}\ \bibnamefont {Tseng}}, \bibinfo {author} {\bibfnamefont {G.}~\bibnamefont {Maslennikov}}, \bibinfo {author} {\bibfnamefont {H.~C.~J.}\ \bibnamefont {Gan}},\ and\ \bibinfo {author} {\bibfnamefont {D.}~\bibnamefont {Matsukevich}},\ }\bibfield  {journal} {\bibinfo  {journal} {arXiv}\ }\href {https://doi.org/10.48550/ARXIV.2104.04168} {10.48550/ARXIV.2104.04168} (\bibinfo {year} {2021}),\ \Eprint {https://arxiv.org/abs/2104.04168} {arXiv:2104.04168 [quant-ph]} \BibitemShut {NoStop}%
\bibitem [{\citenamefont {Chen}\ \emph {et~al.}(2021)\citenamefont {Chen}, \citenamefont {Gan}, \citenamefont {Zhang}, \citenamefont {Matuskevich},\ and\ \citenamefont {Kim}}]{Chen2021}%
  \BibitemOpen
  \bibfield  {author} {\bibinfo {author} {\bibfnamefont {W.}~\bibnamefont {Chen}}, \bibinfo {author} {\bibfnamefont {J.}~\bibnamefont {Gan}}, \bibinfo {author} {\bibfnamefont {J.-N.}\ \bibnamefont {Zhang}}, \bibinfo {author} {\bibfnamefont {D.}~\bibnamefont {Matuskevich}},\ and\ \bibinfo {author} {\bibfnamefont {K.}~\bibnamefont {Kim}},\ }\href {https://doi.org/10.1088/1674-1056/ac01e3} {\bibfield  {journal} {\bibinfo  {journal} {Chin. Phys. B}\ }\textbf {\bibinfo {volume} {30}},\ \bibinfo {pages} {060311} (\bibinfo {year} {2021})}\BibitemShut {NoStop}%
\bibitem [{\citenamefont {Wan}\ \emph {et~al.}(2021)\citenamefont {Wan}, \citenamefont {Choi}, \citenamefont {Kim}, \citenamefont {Shutty},\ and\ \citenamefont {Hayden}}]{Wan2021}%
  \BibitemOpen
  \bibfield  {author} {\bibinfo {author} {\bibfnamefont {K.}~\bibnamefont {Wan}}, \bibinfo {author} {\bibfnamefont {S.}~\bibnamefont {Choi}}, \bibinfo {author} {\bibfnamefont {I.~H.}\ \bibnamefont {Kim}}, \bibinfo {author} {\bibfnamefont {N.}~\bibnamefont {Shutty}},\ and\ \bibinfo {author} {\bibfnamefont {P.}~\bibnamefont {Hayden}},\ }\href {https://doi.org/10.1103/prxquantum.2.040345} {\bibfield  {journal} {\bibinfo  {journal} {{PRX} Quantum}\ }\textbf {\bibinfo {volume} {2}},\ \bibinfo {pages} {040345} (\bibinfo {year} {2021})}\BibitemShut {NoStop}%
\bibitem [{\citenamefont {Bienfait}\ \emph {et~al.}(2019)\citenamefont {Bienfait}, \citenamefont {Satzinger}, \citenamefont {Zhong}, \citenamefont {Chang}, \citenamefont {Chou}, \citenamefont {Conner}, \citenamefont {Dumur}, \citenamefont {Grebel}, \citenamefont {Peairs}, \citenamefont {Povey},\ and\ \citenamefont {Cleland}}]{Bienfait2019}%
  \BibitemOpen
  \bibfield  {author} {\bibinfo {author} {\bibfnamefont {A.}~\bibnamefont {Bienfait}}, \bibinfo {author} {\bibfnamefont {K.~J.}\ \bibnamefont {Satzinger}}, \bibinfo {author} {\bibfnamefont {Y.~P.}\ \bibnamefont {Zhong}}, \bibinfo {author} {\bibfnamefont {H.-S.}\ \bibnamefont {Chang}}, \bibinfo {author} {\bibfnamefont {M.-H.}\ \bibnamefont {Chou}}, \bibinfo {author} {\bibfnamefont {C.~R.}\ \bibnamefont {Conner}}, \bibinfo {author} {\bibfnamefont {{\'{E}}.}~\bibnamefont {Dumur}}, \bibinfo {author} {\bibfnamefont {J.}~\bibnamefont {Grebel}}, \bibinfo {author} {\bibfnamefont {G.~A.}\ \bibnamefont {Peairs}}, \bibinfo {author} {\bibfnamefont {R.~G.}\ \bibnamefont {Povey}},\ and\ \bibinfo {author} {\bibfnamefont {A.~N.}\ \bibnamefont {Cleland}},\ }\href {https://doi.org/10.1126/science.aaw8415} {\bibfield  {journal} {\bibinfo  {journal} {Science}\ }\textbf {\bibinfo {volume} {364}},\ \bibinfo {pages} {368} (\bibinfo {year} {2019})}\BibitemShut {NoStop}%
\bibitem [{\citenamefont {Dumur}\ \emph {et~al.}(2021)\citenamefont {Dumur}, \citenamefont {Satzinger}, \citenamefont {Peairs}, \citenamefont {Chou}, \citenamefont {Bienfait}, \citenamefont {Chang}, \citenamefont {Conner}, \citenamefont {Grebel}, \citenamefont {Povey}, \citenamefont {Zhong},\ and\ \citenamefont {Cleland}}]{Dumur2021}%
  \BibitemOpen
  \bibfield  {author} {\bibinfo {author} {\bibfnamefont {{\'{E}}.}~\bibnamefont {Dumur}}, \bibinfo {author} {\bibfnamefont {K.~J.}\ \bibnamefont {Satzinger}}, \bibinfo {author} {\bibfnamefont {G.~A.}\ \bibnamefont {Peairs}}, \bibinfo {author} {\bibfnamefont {M.-H.}\ \bibnamefont {Chou}}, \bibinfo {author} {\bibfnamefont {A.}~\bibnamefont {Bienfait}}, \bibinfo {author} {\bibfnamefont {H.-S.}\ \bibnamefont {Chang}}, \bibinfo {author} {\bibfnamefont {C.~R.}\ \bibnamefont {Conner}}, \bibinfo {author} {\bibfnamefont {J.}~\bibnamefont {Grebel}}, \bibinfo {author} {\bibfnamefont {R.~G.}\ \bibnamefont {Povey}}, \bibinfo {author} {\bibfnamefont {Y.~P.}\ \bibnamefont {Zhong}},\ and\ \bibinfo {author} {\bibfnamefont {A.~N.}\ \bibnamefont {Cleland}},\ }\href {https://doi.org/10.1038/s41534-021-00511-1} {\bibfield  {journal} {\bibinfo  {journal} {npj Quantum Inf.}\ }\textbf {\bibinfo {volume} {7}},\ \bibinfo {pages} {173} (\bibinfo {year} {2021})}\BibitemShut {NoStop}%
\bibitem [{\citenamefont {Dalvit}\ \emph {et~al.}(2006)\citenamefont {Dalvit}, \citenamefont {de~Matos~Filho},\ and\ \citenamefont {Toscano}}]{Dalvit2006}%
  \BibitemOpen
  \bibfield  {author} {\bibinfo {author} {\bibfnamefont {D.~A.~R.}\ \bibnamefont {Dalvit}}, \bibinfo {author} {\bibfnamefont {R.~L.}\ \bibnamefont {de~Matos~Filho}},\ and\ \bibinfo {author} {\bibfnamefont {F.}~\bibnamefont {Toscano}},\ }\href {https://doi.org/10.1088/1367-2630/8/11/276} {\bibfield  {journal} {\bibinfo  {journal} {New J. Phys.}\ }\textbf {\bibinfo {volume} {8}},\ \bibinfo {pages} {276} (\bibinfo {year} {2006})}\BibitemShut {NoStop}%
\bibitem [{\citenamefont {Sab{\'{\i}}n}\ \emph {et~al.}(2014)\citenamefont {Sab{\'{\i}}n}, \citenamefont {Bruschi}, \citenamefont {Ahmadi},\ and\ \citenamefont {Fuentes}}]{Sabin2014}%
  \BibitemOpen
  \bibfield  {author} {\bibinfo {author} {\bibfnamefont {C.}~\bibnamefont {Sab{\'{\i}}n}}, \bibinfo {author} {\bibfnamefont {D.~E.}\ \bibnamefont {Bruschi}}, \bibinfo {author} {\bibfnamefont {M.}~\bibnamefont {Ahmadi}},\ and\ \bibinfo {author} {\bibfnamefont {I.}~\bibnamefont {Fuentes}},\ }\href {https://doi.org/10.1088/1367-2630/16/8/085003} {\bibfield  {journal} {\bibinfo  {journal} {New J. Phys.}\ }\textbf {\bibinfo {volume} {16}},\ \bibinfo {pages} {085003} (\bibinfo {year} {2014})}\BibitemShut {NoStop}%
\bibitem [{\citenamefont {Hu}\ \emph {et~al.}(2018)\citenamefont {Hu}, \citenamefont {Niu}, \citenamefont {Jin}, \citenamefont {Chen}, \citenamefont {Dong}, \citenamefont {Schmiedmayer},\ and\ \citenamefont {Zhou}}]{Hu2018}%
  \BibitemOpen
  \bibfield  {author} {\bibinfo {author} {\bibfnamefont {D.}~\bibnamefont {Hu}}, \bibinfo {author} {\bibfnamefont {L.}~\bibnamefont {Niu}}, \bibinfo {author} {\bibfnamefont {S.}~\bibnamefont {Jin}}, \bibinfo {author} {\bibfnamefont {X.}~\bibnamefont {Chen}}, \bibinfo {author} {\bibfnamefont {G.}~\bibnamefont {Dong}}, \bibinfo {author} {\bibfnamefont {J.}~\bibnamefont {Schmiedmayer}},\ and\ \bibinfo {author} {\bibfnamefont {X.}~\bibnamefont {Zhou}},\ }\href {https://doi.org/10.1038/s42005-018-0030-7} {\bibfield  {journal} {\bibinfo  {journal} {Commun. Phys.}\ }\textbf {\bibinfo {volume} {1}},\ \bibinfo {pages} {29} (\bibinfo {year} {2018})}\BibitemShut {NoStop}%
\bibitem [{\citenamefont {Wolf}\ \emph {et~al.}(2019)\citenamefont {Wolf}, \citenamefont {Shi}, \citenamefont {Heip}, \citenamefont {Gessner}, \citenamefont {Pezz{\`{e}}}, \citenamefont {Smerzi}, \citenamefont {Schulte}, \citenamefont {Hammerer},\ and\ \citenamefont {Schmidt}}]{Wolf2019}%
  \BibitemOpen
  \bibfield  {author} {\bibinfo {author} {\bibfnamefont {F.}~\bibnamefont {Wolf}}, \bibinfo {author} {\bibfnamefont {C.}~\bibnamefont {Shi}}, \bibinfo {author} {\bibfnamefont {J.~C.}\ \bibnamefont {Heip}}, \bibinfo {author} {\bibfnamefont {M.}~\bibnamefont {Gessner}}, \bibinfo {author} {\bibfnamefont {L.}~\bibnamefont {Pezz{\`{e}}}}, \bibinfo {author} {\bibfnamefont {A.}~\bibnamefont {Smerzi}}, \bibinfo {author} {\bibfnamefont {M.}~\bibnamefont {Schulte}}, \bibinfo {author} {\bibfnamefont {K.}~\bibnamefont {Hammerer}},\ and\ \bibinfo {author} {\bibfnamefont {P.~O.}\ \bibnamefont {Schmidt}},\ }\href {https://doi.org/10.1038/s41467-019-10576-4} {\bibfield  {journal} {\bibinfo  {journal} {Nat. Commun.}\ }\textbf {\bibinfo {volume} {10}},\ \bibinfo {pages} {2929} (\bibinfo {year} {2019})}\BibitemShut {NoStop}%
\bibitem [{\citenamefont {Drechsler}\ \emph {et~al.}(2020)\citenamefont {Drechsler}, \citenamefont {Far{\'{\i}}as}, \citenamefont {Freitas}, \citenamefont {Schmiegelow},\ and\ \citenamefont {Paz}}]{Drechsler2020}%
  \BibitemOpen
  \bibfield  {author} {\bibinfo {author} {\bibfnamefont {M.}~\bibnamefont {Drechsler}}, \bibinfo {author} {\bibfnamefont {M.~B.}\ \bibnamefont {Far{\'{\i}}as}}, \bibinfo {author} {\bibfnamefont {N.}~\bibnamefont {Freitas}}, \bibinfo {author} {\bibfnamefont {C.~T.}\ \bibnamefont {Schmiegelow}},\ and\ \bibinfo {author} {\bibfnamefont {J.~P.}\ \bibnamefont {Paz}},\ }\href {https://doi.org/10.1103/physreva.101.052331} {\bibfield  {journal} {\bibinfo  {journal} {Phys. Rev. A}\ }\textbf {\bibinfo {volume} {101}},\ \bibinfo {pages} {052331} (\bibinfo {year} {2020})}\BibitemShut {NoStop}%
\bibitem [{\citenamefont {Cerrillo}\ and\ \citenamefont {Rodr{\'{\i}}guez}(2021)}]{Cerrillo2021}%
  \BibitemOpen
  \bibfield  {author} {\bibinfo {author} {\bibfnamefont {J.}~\bibnamefont {Cerrillo}}\ and\ \bibinfo {author} {\bibfnamefont {D.}~\bibnamefont {Rodr{\'{\i}}guez}},\ }\href {https://doi.org/10.1209/0295-5075/134/38001} {\bibfield  {journal} {\bibinfo  {journal} {Europhys. Lett.}\ }\textbf {\bibinfo {volume} {134}},\ \bibinfo {pages} {38001} (\bibinfo {year} {2021})}\BibitemShut {NoStop}%
\bibitem [{\citenamefont {Delakouras}\ \emph {et~al.}(2022)\citenamefont {Delakouras}, \citenamefont {Rodr{\'i}guez},\ and\ \citenamefont {Cerrillo}}]{Delakouras2022}%
  \BibitemOpen
  \bibfield  {author} {\bibinfo {author} {\bibfnamefont {A.}~\bibnamefont {Delakouras}}, \bibinfo {author} {\bibfnamefont {D.}~\bibnamefont {Rodr{\'i}guez}},\ and\ \bibinfo {author} {\bibfnamefont {J.}~\bibnamefont {Cerrillo}},\ }\bibfield  {journal} {\bibinfo  {journal} {arXiv}\ }\href {https://doi.org/10.48550/ARXIV.2202.00626} {10.48550/ARXIV.2202.00626} (\bibinfo {year} {2022}),\ \Eprint {https://arxiv.org/abs/2202.00626} {arXiv:2202.00626 [quant-ph]} \BibitemShut {NoStop}%
\bibitem [{\citenamefont {Vahala}\ \emph {et~al.}(2009)\citenamefont {Vahala}, \citenamefont {Herrmann}, \citenamefont {Kn{\"u}nz}, \citenamefont {Batteiger}, \citenamefont {Saathoff}, \citenamefont {H{\"a}nsch},\ and\ \citenamefont {Udem}}]{Vahala2009}%
  \BibitemOpen
  \bibfield  {author} {\bibinfo {author} {\bibfnamefont {K.}~\bibnamefont {Vahala}}, \bibinfo {author} {\bibfnamefont {M.}~\bibnamefont {Herrmann}}, \bibinfo {author} {\bibfnamefont {S.}~\bibnamefont {Kn{\"u}nz}}, \bibinfo {author} {\bibfnamefont {V.}~\bibnamefont {Batteiger}}, \bibinfo {author} {\bibfnamefont {G.}~\bibnamefont {Saathoff}}, \bibinfo {author} {\bibfnamefont {T.~W.}\ \bibnamefont {H{\"a}nsch}},\ and\ \bibinfo {author} {\bibfnamefont {T.}~\bibnamefont {Udem}},\ }\href {https://doi.org/10.1038/nphys1367} {\bibfield  {journal} {\bibinfo  {journal} {Nat. Phys.}\ }\textbf {\bibinfo {volume} {5}},\ \bibinfo {pages} {682} (\bibinfo {year} {2009})}\BibitemShut {NoStop}%
\bibitem [{\citenamefont {Kn{\"u}nz}\ \emph {et~al.}(2010)\citenamefont {Kn{\"u}nz}, \citenamefont {Herrmann}, \citenamefont {Batteiger}, \citenamefont {Saathoff}, \citenamefont {H{\"a}nsch}, \citenamefont {Vahala},\ and\ \citenamefont {Udem}}]{Knuenz2010}%
  \BibitemOpen
  \bibfield  {author} {\bibinfo {author} {\bibfnamefont {S.}~\bibnamefont {Kn{\"u}nz}}, \bibinfo {author} {\bibfnamefont {M.}~\bibnamefont {Herrmann}}, \bibinfo {author} {\bibfnamefont {V.}~\bibnamefont {Batteiger}}, \bibinfo {author} {\bibfnamefont {G.}~\bibnamefont {Saathoff}}, \bibinfo {author} {\bibfnamefont {T.~W.}\ \bibnamefont {H{\"a}nsch}}, \bibinfo {author} {\bibfnamefont {K.}~\bibnamefont {Vahala}},\ and\ \bibinfo {author} {\bibfnamefont {T.}~\bibnamefont {Udem}},\ }\href {https://doi.org/10.1103/PhysRevLett.105.013004} {\bibfield  {journal} {\bibinfo  {journal} {Phys. Rev. Lett.}\ }\textbf {\bibinfo {volume} {105}},\ \bibinfo {pages} {013004} (\bibinfo {year} {2010})}\BibitemShut {NoStop}%
\bibitem [{\citenamefont {Ip}\ \emph {et~al.}(2018)\citenamefont {Ip}, \citenamefont {Ransford}, \citenamefont {Jayich}, \citenamefont {Long}, \citenamefont {Roman},\ and\ \citenamefont {Campbell}}]{Ip2018}%
  \BibitemOpen
  \bibfield  {author} {\bibinfo {author} {\bibfnamefont {M.}~\bibnamefont {Ip}}, \bibinfo {author} {\bibfnamefont {A.}~\bibnamefont {Ransford}}, \bibinfo {author} {\bibfnamefont {A.~M.}\ \bibnamefont {Jayich}}, \bibinfo {author} {\bibfnamefont {X.}~\bibnamefont {Long}}, \bibinfo {author} {\bibfnamefont {C.}~\bibnamefont {Roman}},\ and\ \bibinfo {author} {\bibfnamefont {W.~C.}\ \bibnamefont {Campbell}},\ }\href {https://doi.org/10.1103/PhysRevLett.121.043201} {\bibfield  {journal} {\bibinfo  {journal} {Phys. Rev. Lett.}\ }\textbf {\bibinfo {volume} {121}},\ \bibinfo {pages} {043201} (\bibinfo {year} {2018})}\BibitemShut {NoStop}%
\bibitem [{\citenamefont {Behrle}\ \emph {et~al.}(2023)\citenamefont {Behrle}, \citenamefont {Nguyen}, \citenamefont {Reiter}, \citenamefont {Baur}, \citenamefont {de~Neeve}, \citenamefont {Stadler}, \citenamefont {Marinelli}, \citenamefont {Lancellotti}, \citenamefont {Yelin},\ and\ \citenamefont {Home}}]{Behrle2023}%
  \BibitemOpen
  \bibfield  {author} {\bibinfo {author} {\bibfnamefont {T.}~\bibnamefont {Behrle}}, \bibinfo {author} {\bibfnamefont {T.}~\bibnamefont {Nguyen}}, \bibinfo {author} {\bibfnamefont {F.}~\bibnamefont {Reiter}}, \bibinfo {author} {\bibfnamefont {D.}~\bibnamefont {Baur}}, \bibinfo {author} {\bibfnamefont {B.}~\bibnamefont {de~Neeve}}, \bibinfo {author} {\bibfnamefont {M.}~\bibnamefont {Stadler}}, \bibinfo {author} {\bibfnamefont {M.}~\bibnamefont {Marinelli}}, \bibinfo {author} {\bibfnamefont {F.}~\bibnamefont {Lancellotti}}, \bibinfo {author} {\bibfnamefont {S.}~\bibnamefont {Yelin}},\ and\ \bibinfo {author} {\bibfnamefont {J.}~\bibnamefont {Home}},\ }\href {https://doi.org/10.1103/physrevlett.131.043605} {\bibfield  {journal} {\bibinfo  {journal} {Physical Review Letters}\ }\textbf {\bibinfo {volume} {131}},\ \bibinfo {pages} {043605} (\bibinfo {year} {2023})}\BibitemShut {NoStop}%
\bibitem [{\citenamefont {Kabuss}\ \emph {et~al.}(2012)\citenamefont {Kabuss}, \citenamefont {Carmele}, \citenamefont {Brandes},\ and\ \citenamefont {Knorr}}]{Kabuss2012}%
  \BibitemOpen
  \bibfield  {author} {\bibinfo {author} {\bibfnamefont {J.}~\bibnamefont {Kabuss}}, \bibinfo {author} {\bibfnamefont {A.}~\bibnamefont {Carmele}}, \bibinfo {author} {\bibfnamefont {T.}~\bibnamefont {Brandes}},\ and\ \bibinfo {author} {\bibfnamefont {A.}~\bibnamefont {Knorr}},\ }\href {https://doi.org/10.1103/PhysRevLett.109.054301} {\bibfield  {journal} {\bibinfo  {journal} {Phys. Rev. Lett.}\ }\textbf {\bibinfo {volume} {109}},\ \bibinfo {pages} {054301} (\bibinfo {year} {2012})}\BibitemShut {NoStop}%
\bibitem [{\citenamefont {Kabuss}\ \emph {et~al.}(2013)\citenamefont {Kabuss}, \citenamefont {Carmele},\ and\ \citenamefont {Knorr}}]{Kabuss2013}%
  \BibitemOpen
  \bibfield  {author} {\bibinfo {author} {\bibfnamefont {J.}~\bibnamefont {Kabuss}}, \bibinfo {author} {\bibfnamefont {A.}~\bibnamefont {Carmele}},\ and\ \bibinfo {author} {\bibfnamefont {A.}~\bibnamefont {Knorr}},\ }\href {https://doi.org/10.1103/PhysRevB.88.064305} {\bibfield  {journal} {\bibinfo  {journal} {Phys. Rev. B}\ }\textbf {\bibinfo {volume} {88}},\ \bibinfo {pages} {064305} (\bibinfo {year} {2013})}\BibitemShut {NoStop}%
\bibitem [{\citenamefont {Khaetskii}\ \emph {et~al.}(2013)\citenamefont {Khaetskii}, \citenamefont {Golovach}, \citenamefont {Hu},\ and\ \citenamefont {\v{Z}uti\'{c}}}]{Khaetskii2013}%
  \BibitemOpen
  \bibfield  {author} {\bibinfo {author} {\bibfnamefont {A.}~\bibnamefont {Khaetskii}}, \bibinfo {author} {\bibfnamefont {V.~N.}\ \bibnamefont {Golovach}}, \bibinfo {author} {\bibfnamefont {X.}~\bibnamefont {Hu}},\ and\ \bibinfo {author} {\bibfnamefont {I.}~\bibnamefont {\v{Z}uti\'{c}}},\ }\href {https://doi.org/10.1103/PhysRevLett.111.186601} {\bibfield  {journal} {\bibinfo  {journal} {Phys. Rev. Lett.}\ }\textbf {\bibinfo {volume} {111}},\ \bibinfo {pages} {186601} (\bibinfo {year} {2013})}\BibitemShut {NoStop}%
\bibitem [{\citenamefont {Grudinin}\ \emph {et~al.}(2010)\citenamefont {Grudinin}, \citenamefont {Lee}, \citenamefont {Painter},\ and\ \citenamefont {Vahala}}]{Grudinin2010}%
  \BibitemOpen
  \bibfield  {author} {\bibinfo {author} {\bibfnamefont {I.~S.}\ \bibnamefont {Grudinin}}, \bibinfo {author} {\bibfnamefont {H.}~\bibnamefont {Lee}}, \bibinfo {author} {\bibfnamefont {O.}~\bibnamefont {Painter}},\ and\ \bibinfo {author} {\bibfnamefont {K.~J.}\ \bibnamefont {Vahala}},\ }\href {https://doi.org/10.1103/PhysRevLett.104.083901} {\bibfield  {journal} {\bibinfo  {journal} {Phys. Rev. Lett.}\ }\textbf {\bibinfo {volume} {104}},\ \bibinfo {pages} {083901} (\bibinfo {year} {2010})}\BibitemShut {NoStop}%
\bibitem [{\citenamefont {Beardsley}\ \emph {et~al.}(2010)\citenamefont {Beardsley}, \citenamefont {Akimov}, \citenamefont {Henini},\ and\ \citenamefont {Kent}}]{Beardsley2010}%
  \BibitemOpen
  \bibfield  {author} {\bibinfo {author} {\bibfnamefont {R.~P.}\ \bibnamefont {Beardsley}}, \bibinfo {author} {\bibfnamefont {A.~V.}\ \bibnamefont {Akimov}}, \bibinfo {author} {\bibfnamefont {M.}~\bibnamefont {Henini}},\ and\ \bibinfo {author} {\bibfnamefont {A.~J.}\ \bibnamefont {Kent}},\ }\href {https://doi.org/10.1103/PhysRevLett.104.085501} {\bibfield  {journal} {\bibinfo  {journal} {Phys. Rev. Lett.}\ }\textbf {\bibinfo {volume} {104}},\ \bibinfo {pages} {085501} (\bibinfo {year} {2010})}\BibitemShut {NoStop}%
\bibitem [{\citenamefont {Mahboob}\ \emph {et~al.}(2013)\citenamefont {Mahboob}, \citenamefont {Nishiguchi}, \citenamefont {Fujiwara},\ and\ \citenamefont {Yamaguchi}}]{Mahboob2013}%
  \BibitemOpen
  \bibfield  {author} {\bibinfo {author} {\bibfnamefont {I.}~\bibnamefont {Mahboob}}, \bibinfo {author} {\bibfnamefont {K.}~\bibnamefont {Nishiguchi}}, \bibinfo {author} {\bibfnamefont {A.}~\bibnamefont {Fujiwara}},\ and\ \bibinfo {author} {\bibfnamefont {H.}~\bibnamefont {Yamaguchi}},\ }\href {https://doi.org/10.1103/PhysRevLett.110.127202} {\bibfield  {journal} {\bibinfo  {journal} {Phys. Rev. Lett.}\ }\textbf {\bibinfo {volume} {110}},\ \bibinfo {pages} {127202} (\bibinfo {year} {2013})}\BibitemShut {NoStop}%
\bibitem [{\citenamefont {Kemiktarak}\ \emph {et~al.}(2014)\citenamefont {Kemiktarak}, \citenamefont {Durand}, \citenamefont {Metcalfe},\ and\ \citenamefont {Lawall}}]{Kemiktarak2014}%
  \BibitemOpen
  \bibfield  {author} {\bibinfo {author} {\bibfnamefont {U.}~\bibnamefont {Kemiktarak}}, \bibinfo {author} {\bibfnamefont {M.}~\bibnamefont {Durand}}, \bibinfo {author} {\bibfnamefont {M.}~\bibnamefont {Metcalfe}},\ and\ \bibinfo {author} {\bibfnamefont {J.}~\bibnamefont {Lawall}},\ }\href {https://doi.org/10.1103/PhysRevLett.113.030802} {\bibfield  {journal} {\bibinfo  {journal} {Phys. Rev. Lett.}\ }\textbf {\bibinfo {volume} {113}},\ \bibinfo {pages} {030802} (\bibinfo {year} {2014})}\BibitemShut {NoStop}%
\bibitem [{\citenamefont {Jing}\ \emph {et~al.}(2014)\citenamefont {Jing}, \citenamefont {{\"O}zdemir}, \citenamefont {L{\"u}}, \citenamefont {Zhang}, \citenamefont {Yang},\ and\ \citenamefont {Nori}}]{Jing2014}%
  \BibitemOpen
  \bibfield  {author} {\bibinfo {author} {\bibfnamefont {H.}~\bibnamefont {Jing}}, \bibinfo {author} {\bibfnamefont {S.~K.}\ \bibnamefont {{\"O}zdemir}}, \bibinfo {author} {\bibfnamefont {X.-Y.}\ \bibnamefont {L{\"u}}}, \bibinfo {author} {\bibfnamefont {J.}~\bibnamefont {Zhang}}, \bibinfo {author} {\bibfnamefont {L.}~\bibnamefont {Yang}},\ and\ \bibinfo {author} {\bibfnamefont {F.}~\bibnamefont {Nori}},\ }\href {https://doi.org/10.1103/PhysRevLett.113.053604} {\bibfield  {journal} {\bibinfo  {journal} {Phys. Rev. Lett.}\ }\textbf {\bibinfo {volume} {113}},\ \bibinfo {pages} {053604} (\bibinfo {year} {2014})}\BibitemShut {NoStop}%
\bibitem [{\citenamefont {Zhang}\ \emph {et~al.}(2018{\natexlab{a}})\citenamefont {Zhang}, \citenamefont {Peng}, \citenamefont {{\"O}zdemir}, \citenamefont {Pichler}, \citenamefont {Krimer}, \citenamefont {Zhao}, \citenamefont {Nori}, \citenamefont {Liu}, \citenamefont {Rotter},\ and\ \citenamefont {Yang}}]{Zhang2018}%
  \BibitemOpen
  \bibfield  {author} {\bibinfo {author} {\bibfnamefont {J.}~\bibnamefont {Zhang}}, \bibinfo {author} {\bibfnamefont {B.}~\bibnamefont {Peng}}, \bibinfo {author} {\bibfnamefont {{\c{S}}.~K.}\ \bibnamefont {{\"O}zdemir}}, \bibinfo {author} {\bibfnamefont {K.}~\bibnamefont {Pichler}}, \bibinfo {author} {\bibfnamefont {D.~O.}\ \bibnamefont {Krimer}}, \bibinfo {author} {\bibfnamefont {G.}~\bibnamefont {Zhao}}, \bibinfo {author} {\bibfnamefont {F.}~\bibnamefont {Nori}}, \bibinfo {author} {\bibfnamefont {Y.-x.}\ \bibnamefont {Liu}}, \bibinfo {author} {\bibfnamefont {S.}~\bibnamefont {Rotter}},\ and\ \bibinfo {author} {\bibfnamefont {L.}~\bibnamefont {Yang}},\ }\href {https://doi.org/10.1038/s41566-018-0213-5} {\bibfield  {journal} {\bibinfo  {journal} {Nat. Photonics}\ }\textbf {\bibinfo {volume} {12}},\ \bibinfo {pages} {479} (\bibinfo {year} {2018}{\natexlab{a}})}\BibitemShut {NoStop}%
\bibitem [{\citenamefont {Jiang}\ \emph {et~al.}(2018)\citenamefont {Jiang}, \citenamefont {Maayani}, \citenamefont {Carmon}, \citenamefont {Nori},\ and\ \citenamefont {Jing}}]{Jiang2018}%
  \BibitemOpen
  \bibfield  {author} {\bibinfo {author} {\bibfnamefont {Y.}~\bibnamefont {Jiang}}, \bibinfo {author} {\bibfnamefont {S.}~\bibnamefont {Maayani}}, \bibinfo {author} {\bibfnamefont {T.}~\bibnamefont {Carmon}}, \bibinfo {author} {\bibfnamefont {F.}~\bibnamefont {Nori}},\ and\ \bibinfo {author} {\bibfnamefont {H.}~\bibnamefont {Jing}},\ }\href {https://doi.org/10.1103/PhysRevApplied.10.064037} {\bibfield  {journal} {\bibinfo  {journal} {Phys. Rev. Applied}\ }\textbf {\bibinfo {volume} {10}},\ \bibinfo {pages} {064037} (\bibinfo {year} {2018})}\BibitemShut {NoStop}%
\bibitem [{\citenamefont {Pettit}\ \emph {et~al.}(2019)\citenamefont {Pettit}, \citenamefont {Ge}, \citenamefont {Kumar}, \citenamefont {Luntz-Martin}, \citenamefont {Schultz}, \citenamefont {Neukirch}, \citenamefont {Bhattacharya},\ and\ \citenamefont {Vamivakas}}]{Pettit2019}%
  \BibitemOpen
  \bibfield  {author} {\bibinfo {author} {\bibfnamefont {R.~M.}\ \bibnamefont {Pettit}}, \bibinfo {author} {\bibfnamefont {W.}~\bibnamefont {Ge}}, \bibinfo {author} {\bibfnamefont {P.}~\bibnamefont {Kumar}}, \bibinfo {author} {\bibfnamefont {D.~R.}\ \bibnamefont {Luntz-Martin}}, \bibinfo {author} {\bibfnamefont {J.~T.}\ \bibnamefont {Schultz}}, \bibinfo {author} {\bibfnamefont {L.~P.}\ \bibnamefont {Neukirch}}, \bibinfo {author} {\bibfnamefont {M.}~\bibnamefont {Bhattacharya}},\ and\ \bibinfo {author} {\bibfnamefont {A.~N.}\ \bibnamefont {Vamivakas}},\ }\href {https://doi.org/10.1038/s41566-019-0395-5} {\bibfield  {journal} {\bibinfo  {journal} {Nat. Photonics}\ }\textbf {\bibinfo {volume} {13}},\ \bibinfo {pages} {402} (\bibinfo {year} {2019})}\BibitemShut {NoStop}%
\bibitem [{\citenamefont {Sheng}\ \emph {et~al.}(2020)\citenamefont {Sheng}, \citenamefont {Wei}, \citenamefont {Yang},\ and\ \citenamefont {Wu}}]{Sheng2020}%
  \BibitemOpen
  \bibfield  {author} {\bibinfo {author} {\bibfnamefont {J.}~\bibnamefont {Sheng}}, \bibinfo {author} {\bibfnamefont {X.}~\bibnamefont {Wei}}, \bibinfo {author} {\bibfnamefont {C.}~\bibnamefont {Yang}},\ and\ \bibinfo {author} {\bibfnamefont {H.}~\bibnamefont {Wu}},\ }\href {https://doi.org/10.1103/PhysRevLett.124.053604} {\bibfield  {journal} {\bibinfo  {journal} {Phys. Rev. Lett.}\ }\textbf {\bibinfo {volume} {124}},\ \bibinfo {pages} {053604} (\bibinfo {year} {2020})}\BibitemShut {NoStop}%
\bibitem [{\citenamefont {Khurgin}\ \emph {et~al.}(2012)\citenamefont {Khurgin}, \citenamefont {Pruessner}, \citenamefont {Stievater},\ and\ \citenamefont {Rabinovich}}]{Khurgin2012}%
  \BibitemOpen
  \bibfield  {author} {\bibinfo {author} {\bibfnamefont {J.~B.}\ \bibnamefont {Khurgin}}, \bibinfo {author} {\bibfnamefont {M.~W.}\ \bibnamefont {Pruessner}}, \bibinfo {author} {\bibfnamefont {T.~H.}\ \bibnamefont {Stievater}},\ and\ \bibinfo {author} {\bibfnamefont {W.~S.}\ \bibnamefont {Rabinovich}},\ }\href {https://doi.org/10.1103/PhysRevLett.108.223904} {\bibfield  {journal} {\bibinfo  {journal} {Phys. Rev. Lett.}\ }\textbf {\bibinfo {volume} {108}},\ \bibinfo {pages} {223904} (\bibinfo {year} {2012})}\BibitemShut {NoStop}%
\bibitem [{\citenamefont {Zhang}\ \emph {et~al.}(2018{\natexlab{b}})\citenamefont {Zhang}, \citenamefont {Zou}, \citenamefont {Yang}, \citenamefont {Jing}, \citenamefont {Dong}, \citenamefont {Guo},\ and\ \citenamefont {Zou}}]{Zhang2018a}%
  \BibitemOpen
  \bibfield  {author} {\bibinfo {author} {\bibfnamefont {Y.-L.}\ \bibnamefont {Zhang}}, \bibinfo {author} {\bibfnamefont {C.-L.}\ \bibnamefont {Zou}}, \bibinfo {author} {\bibfnamefont {C.-S.}\ \bibnamefont {Yang}}, \bibinfo {author} {\bibfnamefont {H.}~\bibnamefont {Jing}}, \bibinfo {author} {\bibfnamefont {C.-H.}\ \bibnamefont {Dong}}, \bibinfo {author} {\bibfnamefont {G.-C.}\ \bibnamefont {Guo}},\ and\ \bibinfo {author} {\bibfnamefont {X.-B.}\ \bibnamefont {Zou}},\ }\href {https://doi.org/10.1088/1367-2630/aadc9f} {\bibfield  {journal} {\bibinfo  {journal} {New J. Phys.}\ }\textbf {\bibinfo {volume} {20}},\ \bibinfo {pages} {093005} (\bibinfo {year} {2018}{\natexlab{b}})}\BibitemShut {NoStop}%
\bibitem [{\citenamefont {Navarrete-Benlloch}\ \emph {et~al.}(2014)\citenamefont {Navarrete-Benlloch}, \citenamefont {Garc{\'i}a-Ripoll},\ and\ \citenamefont {Porras}}]{NavarreteBenlloch2014}%
  \BibitemOpen
  \bibfield  {author} {\bibinfo {author} {\bibfnamefont {C.}~\bibnamefont {Navarrete-Benlloch}}, \bibinfo {author} {\bibfnamefont {J.~J.}\ \bibnamefont {Garc{\'i}a-Ripoll}},\ and\ \bibinfo {author} {\bibfnamefont {D.}~\bibnamefont {Porras}},\ }\href {https://doi.org/10.1103/physrevlett.113.193601} {\bibfield  {journal} {\bibinfo  {journal} {Physical Review Letters}\ }\textbf {\bibinfo {volume} {113}},\ \bibinfo {pages} {193601} (\bibinfo {year} {2014})}\BibitemShut {NoStop}%
\bibitem [{\citenamefont {{S{\'a}nchez Mu{\~n}oz}}\ and\ \citenamefont {Jaksch}(2021)}]{SanchezMunoz2021}%
  \BibitemOpen
  \bibfield  {author} {\bibinfo {author} {\bibfnamefont {C.}~\bibnamefont {{S{\'a}nchez Mu{\~n}oz}}}\ and\ \bibinfo {author} {\bibfnamefont {D.}~\bibnamefont {Jaksch}},\ }\href {https://doi.org/10.1103/physrevlett.127.183603} {\bibfield  {journal} {\bibinfo  {journal} {Physical Review Letters}\ }\textbf {\bibinfo {volume} {127}},\ \bibinfo {pages} {183603} (\bibinfo {year} {2021})}\BibitemShut {NoStop}%
\bibitem [{\citenamefont {Cirac}\ \emph {et~al.}(1993)\citenamefont {Cirac}, \citenamefont {Parkins}, \citenamefont {Blatt},\ and\ \citenamefont {Zoller}}]{Cirac1993}%
  \BibitemOpen
  \bibfield  {author} {\bibinfo {author} {\bibfnamefont {J.~I.}\ \bibnamefont {Cirac}}, \bibinfo {author} {\bibfnamefont {A.~S.}\ \bibnamefont {Parkins}}, \bibinfo {author} {\bibfnamefont {R.}~\bibnamefont {Blatt}},\ and\ \bibinfo {author} {\bibfnamefont {P.}~\bibnamefont {Zoller}},\ }\href {https://doi.org/10.1103/physrevlett.70.556} {\bibfield  {journal} {\bibinfo  {journal} {Physical Review Letters}\ }\textbf {\bibinfo {volume} {70}},\ \bibinfo {pages} {556} (\bibinfo {year} {1993})}\BibitemShut {NoStop}%
\end{thebibliography}%

\end{document}